\documentstyle[12pt,a4,epsf]{article}

\newcommand{\news}{\setcounter{equation}{0}}
\renewcommand{\theequation}{\thesection.\arabic{equation}}
\def\eqn{\begin{equation}}
\def\eeqn{\end{equation}}
\def\arr{\begin{array}}
\def\earr{\end{array}}
\def\eqna{\begin{eqnarray}}
\def\eeqna{\end{eqnarray}}
\def\a{\alpha}
\def\b{\beta}
\def\s{\sigma}
\def\d{\delta}

\def\O{\Omega}
\def\e{\epsilon}
\def\th{\theta}
\def\m{\mu}
\def\la{\lambda}

\def\t{\tau}
\def\p{\partial}

\def\ha{\hat{\alpha}}
\def\hb{\hat{\beta}}
\def\hg{\hat{\gamma}}
\def\hth{\hat{\theta}}
\font\mybb=msbm10 at 12pt
\def\bb#1{\hbox{\mybb#1}}

\def\bE {\bb{E}}

\begin{document}

\vspace*{-.6in}
\thispagestyle{empty}
\begin{flushright}
DAMTP-R/97/63\\
\end{flushright}

{\Large
\begin{center}
{\bf Torons and black hole entropy}
\end{center}}
\vspace{.3in}
\begin{center}
Miguel S. Costa\footnote{M.S.Costa@damtp.cam.ac.uk}
and Malcolm J. Perry\footnote{malcolm@damtp.cam.ac.uk}\\
\vspace{.1in}
\emph{D.A.M.T.P.\\ University of Cambridge \\ Cambridge CB3 9EW \\ UK}
\end{center}

\vspace{.5in}

\begin{abstract}
We consider a supersymmetric system of D-5-branes compactified on
$T^4\times S^1$ with a self-dual
background field strength on $T^4$ and carrying left-moving momentum
along $S^1$. The corresponding supergravity
solution describes a 5-dimensional black hole with a regular
horizon. The entropy of this black hole may be explained in terms
of the Landau
degeneracy for open strings stretching between different branes. In
the gauge theory approximation this D-5-brane system is described by a
super Yang-Mills theory with a t'Hooft twist. By choosing a
supersymmetric branch of the theory we obtain perfect agreement with
the entropy formula. The result relies on the number of massless
torons associated with the gauge field components that obey twisted
boundary conditions.
\end{abstract}
\newpage

\section{Introduction}
\news
One of the most attractive features of superstring theory as an
unifying theory is that it naturally incorporates gravity and gauge
theories. In other words, the dynamics of the background spacetime are
determined by a given supergravity theory, and the dynamics for the
massless modes associated with the solitonic objects (D-branes
\cite{Polc,Polc1,Witt1}, solitonic 5-brane \cite{Call..} and intersecting
configurations \cite{Doug,Doug1}) are determined by a
given worldvolume gauge theory. A remarkable example of the interplay
between these different aspects of superstring theory is the
explanation of the statistical origin of the Bekenstein-Hawking
entropy for supersymmetric black holes with a regular horizon and
carrying Ramond-Ramond charge
\cite{StroVafa,CallMald,Brec..1,MaldStro,John..,Mald}. These
black holes may be studied in the low string coupling limit  by
identifying them with an excited D-brane system. The degeneracy of BPS
states is then given by the number of massless open string
states that reproduce the excited D-brane system. This argument,
however, requires an assumption on which open string states are the
relevant ones for the black hole entropy counting \cite{CallMald}. A
more precise
treatment involves considering the supersymmetric gauge theory
describing the D-brane system, taking into account the interactions
between different massless modes on the branes \cite{Mald}. 

In a recent paper \cite{CostaPerry}, we have addressed the problem of
black hole entropy counting for a supersymmetric D-brane
system with magnetic condensates on the constituent D-branes'
worldvolume theory. The agreement with the semi-classical entropy formula was
shown by considering the Landau degeneracy for open strings stretching
between different branes \cite{Abou..}. In this paper we will explore
the existence
of these Landau levels from the gauge theory point of view. In
particular, we shall see that the Landau levels arise within the gauge
theory context as torons, i.e. instantons on a torus. The
matching with the semi-classical entropy formula uses
non-trivial results about the dimension of the space of
$\Theta$-functions on $T^4$ \cite{Igusa,Hano}. Also, in the gauge theory
approximation we have control on 
the interactions between the different massless fields on the
D-branes, and no assumption regarding which string states contribute
to the entropy is necessary. 

Our work provides another example of the interplay between string
theory, black-hole physics and gauge theories. We will use old
results of gauge theories on compact spaces with twisted boundary
conditions on the fields \cite{tHoo,tHoo1,Baal,Baal1}. This work
constituted an early attempt to understand quark
confinement and has been placed in the context of D-brane physics to
describe the worldvolume theory of D-branes carrying magnetic
fluxes \cite{GuraRamg,HashTayl}. 

We start by describing the D-brane system studied in this paper. We
will consider a system of coincident D-5-branes
on $T^4\times S^1$ associated with a 5-dimensional black hole. Each
D-5-brane will have a constant (anti)self-dual field strength on
$T^4$ and will carry momentum along $S^1$. The corresponding
gauge theory describing the massless modes of this D-brane system is
given by the compactification of $D=10$ super Yang-Mills theory to 6
dimensions \cite{Witt1} (we will ignore the Born-Infeld corrections to the
action \cite{Leigh}). To derive our D-brane system we start with
the supersymmetric
configuration of D-2-branes on $T^4$ intersecting at $SU(2)$
angles \cite{Berk..}. We consider $N_i$ D-2-branes, which will be
called of type $i$, and place them with respect to the coordinate
system $x^2,...x^5$ on $T^4$ according to (following the notation of
\cite{CostaPerry})
\eqn
\arr{rl}
2_i:&X^2,X^4,\\
    &\phi^3_{(i)}=\tan{\th_i}\ X^2\ ,\\
    &\phi^5_{(i)}=\pm\tan{\th_i}\ X^4\ ,
\earr
\label{1.1}
\eeqn
for $i=1,...n$, i.e. we have $n$ different types of D-2-branes. The
angles $\theta_i$ obey the quantisation conditions
\eqn
\tan{\th_i}=\frac{q_i}{p_i}\frac{R_3}{R_2}=
\frac{\bar{q}_i}{\bar{p}_i}\frac{R_5}{R_4}\ .
\label{1.2}
\eeqn
In other words, the type $i$ D-2-branes are wrapped on the $(p_i,q_i)$ and
$(\bar{p}_i,\bar{q}_i)$ cycles of $T^4=T^2\times T^2$. 

We proceed by performing T-duality transformations along the $x^1,x^3$-
and $x^5$-directions. The resulting type $i$ D-5-branes are described by
\eqna
5_i:&&X^1,X^2,X^3,X^4,X^5,\nonumber\\
    &&2\pi\a'G_{23}^{(i)}=\tan{\th_i}=
     \frac{q_i}{p_i}\frac{\a'}{R_2R_3}\ ,\label{1.3}\\
    &&2\pi\a'G_{45}^{(i)}=\pm\tan{\th_i}=
     \pm\frac{\bar{q}_i}{\bar{p}_i}\frac{\a'}{R_4R_5} \ ,\nonumber
\eeqna
where $G_{\a\b}^{(i)}$ is the field strength of the
type $i$ D-5-branes. It is (anti)self-dual on the subspace $T^4$. For
each $i$ we have $N_i$ D-5-branes winding $p_i$ and
$\bar{p}_i$ times in the $x^2$- and $x^4$-directions,
respectively. They also carry $N_iq_i\bar{p}_i$ and $N_i\bar{q}_ip_i$
units of D-3-brane charge, as well as $N_iq_i\bar{q}_i$
units of D-string charge. This follows from the fact that each
type $i$ D-5-brane carries fluxes $2\pi q_i\bar{p}_i$ and
$2\pi\bar{q}_ip_i$ in the $x^2x^3$- and $x^4x^5$ 2-torus,
respectively. We may now
excite the D-brane system while preserving some supersymmetry by
allowing these D-branes to carry all together $N$ units of left-moving
momentum along the $x^1$-direction, 
\eqn
P=\frac{N}{R_1}\ .
\label{1.4}
\eeqn
Our configuration preserves $1/8$ of the spacetime
supersymmetry and therefore $1/4$ of the D-5-branes' worldvolume
supersymmetry. Through this paper we will just consider the self-dual
field strength case. The analysis for the antiself-dual
case is identical.

We begin in section 2 by writing the supergravity solution that is
associated with our D-brane system. We shall then derive the
Bekenstein-Hawking entropy formula in terms of quantised charges. In
section 3 we shall see how this entropy arises from the Landau
degeneracy for strings stretching between different branes. We will be
brief since the results are essentially the same as those presented in
\cite{CostaPerry}. In section 4 we shall study the gauge theory
describing our D-brane system. The main result will be that there are
several massless torons whose degeneracy exactly match the entropy
formula. This matching arises due to the fact that these torons are
described in terms of $\Theta$-functions on $T^4$. These functions form
a complex linear space whose dimension equals the number of
Landau levels.

\section{Supergravity solution}
\news

The supergravity solution describing the long range fields of the
D-5-brane system discussed in the introduction may be obtained by
following essentially the same steps. We start with the supergravity
solution describing $n$ D-branes intersecting at $SU(2)$ angles
\cite{Brec..2,Bala..} and use the supergravity T-duality rules
\cite{Berg..}. The
resulting metric and dilaton field are
\eqna
ds^{2}&=& H^{\frac{1}{2}}\left[ H^{-1}
\left( -dt^2+dx_1^2+\frac{\a}{r^2}(dx_1-dt)^2\right)\right.
\nonumber\\\nonumber\\
&&\left. +\tilde{H}^{-1}_{\phantom{\frac{1}{1}}}ds^2(T^4)+
ds^2(\bE^4)\right]\ ,
\label{2.1}\\\nonumber\\
e^{2(\phi-\phi_{\infty})}&=&H\tilde{H}^{-2}\ ,\nonumber
\eeqna
where
\eqna
H&=&1+\sum_{i}\frac{\m_i}{r^2}+
\sum_{i<j}\frac{\m_i\m_j}{r^4}\sin^2{(\th_i-\th_j)}\ ,
\nonumber\\\label{2.2}\\
\tilde{H}&=&1+\sum_{i}\frac{\m_i}{r^2}\cos^2{\th_i}\ ,\nonumber
\eeqna
with $r$ the radial coordinate on the 4-dimensional Euclidean space
$\bE^4$. The $i$ and $j$ indices in the sums run from 1 to $n$.
The constants $\mu_i$ determine the ADM mass of the
D-5-branes, the angles $\th_i$
determine the D-5-branes' field strength and the constant $\a$ the
left-moving momentum carried by all the D-5-branes along the
$x^1$-direction. If $\th_i=0$ for all $i$ we obtain the D-5-brane solution
and if $\th_i=\pi/2$ the D-string solution. If $\th_i=\th_j$ for all
$i$ and $j$ we have enhancement of supersymmetry as our solution
breaks $1/2^2$ of the spacetime supersymmetry (1/2 due to the
D-5-branes and the other 1/2 due to the momentum modes along the
$x^1$-direction). From the worldvolume gauge
theory perspective this fact follows because the non-vanishing
worldvolume field strength is on the $U(1)$ center of the gauge group
and the non-linear realization of the worldvolume supersymmetry may be
used to cancel the variation of the gaugino field under
worldvolume supersymmetry transformations \cite{HarvMoore}. In this
case the horizon area vanishes. Note also that for $n=2$ with $\th_1=0$
and $\th_2=\pi/2$ we recover the D-5-brane/D-string configuration
used by Callan and Maldacena \cite{CallMald}.

The solution (\ref{2.1}) may be reduced to five-dimensions giving a
black hole solution with a regular horizon. From the area of the
horizon we can determine the Bekenstein-Hawking entropy
\eqn
S_{BH}=\frac{A_H}{4G_N^{(5)}}=\frac{A_3}{4G_N^{(5)}}
\sqrt{\a\left(\sum_{i<j}\mu_i\mu_j\sin^2{(\th_i-\th_j)}\right)}\ ,
\label{2.3}
\eeqn
where $G_N^{(5)}$ is the five-dimensional Newton constant and $A_3$
is the unit 3-sphere volume. 

Next, we want to write the
black hole entropy in terms of the microscopic quantities that define
our D-brane system \cite{CallMald}. The mass of $N_i$ D-5-branes
defined by (\ref{1.3}) is \cite{Polc1}
\eqn
M_{5_iD}=N_i\frac{R_1}{g\a'^3}
\sqrt{\left( p_iR_2R_3\right)^2 + \left( q_i\a'\right)^2}
\sqrt{\left( \bar{p}_iR_4R_5\right)^2 +
\left( \bar{q}_i\a'\right)^2}\ ,
\label{2.4}
\eeqn
while the corresponding ADM mass that is obtained from the compactified
supergravity solution is
\eqn
M_{5_iD}=2\mu_i\left(\frac{16\pi G_N^{(5)}}{A_3}\right)^{-1}\ .
\label{2.5}
\eeqn
The left-moving momentum carried by all the D-5-branes may also be related
to the constant $\a$ in the solution (\ref{2.1}) by
\eqn
\a=\frac{4G_N^{(5)}N}{\pi R_1}\ .
\label{2.6}
\eeqn
Using the conditions in (\ref{1.3}) for the angles $\th_i$ and writing
the Newton coupling constant in terms of string theory quantities we
obtain the following expression for the black hole entropy
\eqn
S=2\pi\sqrt{N\sum_{i<j}^{\phantom{1}}N_iN_jn_L^{ij}\bar{n}_L^{ij}}\ ,
\label{2.7}
\eeqn
where
\eqn
n_L^{ij}=|p_jq_i-p_iq_j|\ ,\ \ \ 
\bar{n}_L^{ij}=|\bar{p}_j\bar{q}_i-\bar{p}_i\bar{q}_j|\ .
\label{2.8}
\eeqn
The numbers $n_L^{ij}$ and $\bar{n}_L^{ij}$ appear in the entropy
formula because of the existence of Landau levels in the $x^2,x^3$- and
$x^4,x^5$-directions for open strings with ends on the type $i$ and $j$
D-5-branes. In the gauge theory picture they will correspond to the
number of massless torons associated to such pair of D-5-branes.

\section{String Theory description}
\news

In this section we derive the entropy formula (\ref{2.7}) by studying
the excitations of open strings ending on the D-5-branes. The
analysis entirely parallels that of \cite{CostaPerry}, therefore we
will be brief.

The open strings describing the excitations of our D-brane system may
be divided in two main sectors. The first one is associated with
strings with both ends on the type $i$ D-5-branes. The corresponding
bosonic massless modes are associated with a gauge field $A_{\a}$ $(\a
=0,...,5)$ and 4 scalar fields $\phi_m$ $(m=1,...,4)$. We shall assume
that these excitations do not contribute to the entropy formula. This
assumption will be justified in the next section by using the
gauge theory description of our D-brane system since the interactions
between massless fields on the branes are taken into account.

The second sector is associated with open strings with ends on
different type $i$ and $j$ D-5-branes. The corresponding massless
modes are described by torons on $T^4$ and will be responsible for the
entropy of the system. These strings obey Neumann boundary conditions
at both ends in the $x^0,x^1$-directions and Dirichlet boundary
conditions at both ends in the transverse
$x^6,x^7,x^8,x^9$-directions. Defining the complex world-sheet scalar
fields $Z_k=X^{2k}+iX^{2k+1}$ for $k=1,2$, the boundary conditions along
these directions are
\eqn
\arr{ll}
\p_{\s}Z_k=-i\tan{\th_i}\ \p_{\t}Z_k\ ,\ \ \ &\s =0,\\
\p_{\s}Z_k=-i\tan{\th_j}\ \p_{\t}Z_k\ ,&\s =\pi.
\earr
\label{3.1}
\eeqn
The corresponding mode expansion is \cite{Abou..}
\eqn
Z_k=z_k+i\left[ \sum_{n=1}^{\infty}a^{k}_{n-\e}\phi_{n-\e}(\t ,\s )
-\sum_{n=0}^{\infty}a_{n+\e}^{k\dagger}\phi_{-n-\e}(\t ,\s )\right]\ ,
\label{3.2}
\eeqn
where
\eqn
\phi_{n-\e} =
\frac{1}{\sqrt{|n-\e|}}\cos{[(n-\e)\s+\th_i]}e^{-i(n-\e)\t}\ ,
\label{3.3}
\eeqn
with $\e=(\th_i-\th_j)/\pi$ and $n$ an integer. Note that without loss
of generality we may assume that $\e$ lies between 0 and 1. The real
operators $a^{k}_{n-\e}$, $a^{k}_{n+\e}$ and $a_{n-\e}^{k\dagger}$,
$a_{n+\e}^{k\dagger}$ are annihilation and creation operators,
respectively.

The zero modes $z_k$ and $z_k^{\dagger}$ are non-commuting
variables. In terms of the $x^{\hat{\a}}$-coordinates
$(\hat{\a}=2,3,4,5)$ the corresponding non-vanishing commutators are
\eqn
\left[ x^2,x^3 \right] =\left[ x^4,x^5 \right] = 
i\frac{\pi}{\tan{\th_i}-\tan{\th_j}}\ .
\label{3.4}
\eeqn
As a consequence, the background gauge fields obey a
Dirac quantisation condition (which follows from the flux
quantisation), and for an appropriate choice the zero modes
$x^2$ and $x^4$ can only take the values
\eqna
x^2_r=\frac{r}{n_L^{ij}}L\ ,\ \ \ &L=p_ip_jR_2\ ,
\nonumber\\\label{3.5}\\
x^4_{\bar{r}}=\frac{\bar{r}}{\bar{n}_L^{ij}}\bar{L}\ ,\ \ \ &
\bar{L}=\bar{p}_i\bar{p}_jR_4\ ,\nonumber
\eeqna
where $n_L^{ij}$ and $\bar{n}_L^{ij}$ are given in (\ref{2.8}) and we
have assumed that $p_i$ and $p_j$ ($\bar{p}_i$ and $\bar{p}_j$) are
co-prime. If this is the case, the system has periodicity $L$
and $\bar{L}$ along the $x^2$- and $x^4$-directions, respectively. Thus,
the degeneracy of any string state is $n_L^{ij}\bar{n}_L^{ij}$,
i.e. the string spectrum is found by acting with creation operators on
the degenerated ground states
\eqn
\arr{ll}
|x^2_r,x^4_{\bar{r}}\rangle\ ,\ \ \ \ &
r=1,...,n_L^{ij}\ ,\\
&\bar{r}=1,...,\bar{n}_L^{ij}\ .
\earr
\label{3.6}
\eeqn
If $p_i$ and $p_j$ ($\bar{p}_i$ and $\bar{p}_j$) are not co-prime,
i.e. if there is an integer $l$ ($\bar{l}$) such that $p_i=lp'_i$ and
$p_j=lp'_j$ ($\bar{p}_i=\bar{l}\bar{p}'_i$ and 
$\bar{p}_j=\bar{l}\bar{p}'_j$) with $p'_i$ and $p'_j$ ($\bar{p}'_i$
and $\bar{p}'_j$) co-prime, then the zero modes $x^2$ and $x^4$ can
only take the values (see \cite{CostaPerry} for details)
\eqna
x^2_r=\frac{r}{n_L'^{ij}}L'\ ,\ \ \ &L'=lp'_ip'_jR_2\ ,
\nonumber\\\label{3.7}\\
x^4_{\bar{r}}=\frac{\bar{r}}{\bar{n}_L'^{ij}}\bar{L'}\ ,\ \ \ &
\bar{L'}=\bar{l}\bar{p'}_i\bar{p'}_jR_4\ ,\nonumber
\eeqna
where
\eqn
n_L'^{ij}=|p'_jq_i-p'_iq_j|\ ,\ \ \ 
\bar{n}_L'^{ij}=|\bar{p'}_j\bar{q}_i-\bar{p'}_i\bar{q}_j|\ .
\label{3.8}
\eeqn
In this case there are $n_L'^{ij}\bar{n}_L'^{ij}$ Landau levels but
each one is itself $l\bar{l}$ times degenerated. Thus, the ground
states
\eqn
\arr{lll}
|x^2_{r,s},x^4_{\bar{r},\bar{s}}\rangle\ ,\ \ \ 
&r=1,...,n_L^{ij}\ ,\ &s=1,...,l\ ,\\
&\bar{r}=1,...,\bar{n}_L^{ij}\ ,&\bar{s}=1,...,\bar{l}\ ,
\earr
\label{3.9}
\eeqn
again have degeneracy $n_L^{ij}\bar{n}_L^{ij}$.

In order to find the massless modes associated with the previous open
strings we have to consider the worldsheet fermionic fields. The
spectrum is essentially similar to the one presented in \cite{Berk..}
as we are considering the T-dual configuration. We shall use the
notation of \cite{CostaPerry} for the worldsheet fields mode
expansion. Consider first the NS sector of the theory. The zero energy
is $E_0=-\frac{1}{2}+\e$, where $\e$ was defined in (\ref{3.2}). The
spectrum of the low-lying bosonic states is given by
\eqna
\Psi^{k\dagger}_{\frac{1}{2}-\e}
\left(a^{1\dagger}_{\e}\right)^{m_1}
\left(a^{2\dagger}_{\e}\right)^{m_2}|x^2_r,x^4_{\bar{r}}\rangle\ ,
&\ \ \ \ \ \ \a'M^2=\e(m_1+m_2)\ ,
\label{3.10}\\\nonumber\\
\Psi^{k\dagger}_{\frac{1}{2}+\e}
\left(a^{1\dagger}_{\e}\right)^{m_1}
\left(a^{2\dagger}_{\e}\right)^{m_2}|x^2_r,x^4_{\bar{r}}\rangle\ ,
&\ \ \ \a'M^2=\e(2+m_1+m_2)\ ,
\label{3.11}\\\nonumber\\
\Psi^{m\dagger}_{\frac{1}{2}}
\left(a^{1\dagger}_{\e}\right)^{m_1}
\left(a^{2\dagger}_{\e}\right)^{m_2}|x^2_r,x^4_{\bar{r}}\rangle\ ,
&\ \ \ \a'M^2=\e(1+m_1+m_2)\ ,
\label{3.12}
\eeqna
where the states $|x^2_r,x^4_{\bar{r}}\rangle$ correspond to the
Landau levels (\ref{3.6}) (or (\ref{3.9}) if that is the case). The
real fermionic creation operators $\Psi^{k\dagger}_{\frac{1}{2}-\e}$ and
$\Psi^{k\dagger}_{\frac{1}{2}+\e}$ for $k=1,2$ correspond to the two
complex fermionic worldsheet fields in the $T^4$ directions. The
fermionic creation operators $\Psi^{m\dagger}_{\frac{1}{2}}$ for
$m=1,...,4$ correspond to the four transverse directions to the
D-5-branes. This spectrum should be doubled because strings are
oriented. The massless states in the NS sector are obtained by taking
$m_1=m_2=0$ in the states (\ref{3.10}). These contribute with
precisely $4n^{ij}_L\bar{n}^{ij}_L$ massless bosonic excitations for
each pair of D-5-branes of type $i$ and $j$. 

We now turn to the Ramond sector of the theory. The corresponding
massless states form a representation of the 6-dimensional Dirac
algebra, transforming as a spinor under $SO(1,5)$. A basis for the
vacuum states is
\eqn
|s_1,s_2,s_3,x^2_r,x^4_{\bar{r}}\rangle\ ,
\label{3.13}
\eeqn
where $s_i=\pm$. The GSO projection removes half of the states. The
fact that physical states are annihilated by the zero mode of the
supersymmetry generator removes a further half of these states. In the
frame $p_0=p_1$ the latter condition becomes $s_1=+$. The vacuum
states may then be taken to be
\eqn
|+,\pm,\pm,x^2_r,x^4_{\bar{r}}\rangle\ ,
\label{3.14}
\eeqn
Considering both orientations of the strings we end up with
$4n^{ij}_L\bar{n}^{ij}_L$ massless fermionic excitations for each pair
of D-5-branes of type $i$ and $j$. The fermionic partners of the
massive tower of states (3.10-12) are then given by
\eqna
\left(a^{1\dagger}_{\e}\right)^{m_1}
\left(a^{2\dagger}_{\e}\right)^{m_2}
|+,\pm,\pm,x^2_r,x^4_{\bar{r}}\rangle\ ,
&\ \ \ \ \a'M^2=\e(m_1+m_2)\ ,
\label{3.15}\\\nonumber\\
\Psi^{1\dagger}_{\e}\Psi^{2\dagger}_{\e}
\left(a^{1\dagger}_{\e}\right)^{m_1}
\left(a^{2\dagger}_{\e}\right)^{m_2}
|+,\pm,\pm,x^2_r,x^4_{\bar{r}}\rangle\ ,
&\a'M^2=\e(2+m_1+m_2)\ ,
\label{3.16}\\\nonumber\\
\Psi^{k\dagger}_{\e}
\left(a^{1\dagger}_{\e}\right)^{m_1}
\left(a^{2\dagger}_{\e}\right)^{m_2}
|+,\pm,\mp,x^2_r,x^4_{\bar{r}}\rangle\ ,
&\a'M^2=\e(1+m_1+m_2)\ .
\label{3.17}
\eeqna

To count the entropy associated with the excited D-brane system
representing our black hole we just have to realize that we have a gas
of $4\sum_{i<j}N_iN_jn_L^{ij}\bar{n}_L^{ij}$ massless bosonic and fermionic
species carrying all together the left-moving momentum
$P=\frac{N}{R_1}$. For $N>>\sum_{i<j}N_iN_jn_L^{ij}\bar{n}_L^{ij}$
the corresponding entropy is given by the formula (\ref{2.7}). If the
previous condition does not hold our black hole is represented by a
single D-5-brane of each type wrapped $N_i$ times around the
$x^1$-direction \cite{MaldSuss}. For $n=2$, i.e. for a system with
just two D-5-branes, we 
have a gas with $4n_L^{12}\bar{n}_L^{12}$ bosonic and fermionic
species and the momentum carried by these strings is quantised in
units of $(N_1N_2R_1)^{-1}$. To carry the left-moving momentum
$P=\frac{N}{R_1}$ the system is at the level $N'=N_1N_2N$ and we
obtain the correct result for $NN_1N_2>>n_L^{12}\bar{n}_L^{12}$.
For $n>2$, the N units of momentum along the
$x^1$-direction have to be distributed according to
\eqn
\frac{N}{R}=\sum_{i<j}\left(\frac{1}{N_iN_jR_1}\sum_{k}kn_k\right)\ ,
\label{3.18}
\eeqn
where the last sum is over the $4n_L^{ij}\bar{n}_L^{ij}$ bosonic and
fermionic species, $k$ is the number of quanta of momentum and $n_k$
the corresponding
occupation number. We expect the entropy formula (\ref{2.7}) to remain
valid in the limit $NN_iN_j>>n_L^{ij}\bar{n}_L^{ij}$.

\section{Gauge Theory description}
\news

In this section we will study the gauge theory describing the massless
modes that live on our D-brane system. We shall consider the case
$N_i=1$ for every $i=1,...,n$. The generalisation for arbitrary $N_i$,
either for $N_i$ D-5-branes of the type $i$ on top of each other, or for a
single D-5-brane of each type wrapping $N_i$ times around the
$x^1$-direction, is straightforward. We will comment on this later.

The D-brane dynamics is determined by the Born-Infeld action
\cite{Leigh} and the corresponding, and still not well understood,
non-abelian generalisation \cite{Tsey}. We will work in the linear
approximation where the action describing a system of D-5-branes is
given by the dimensional reduction to 6 dimensions of the $D=10$
$U(N)$ super Yang Mills action ($N$ is still to be
specified). Supersymmetry of our
field configuration guarantees that the number of massless particles
remains unchanged in both descriptions (although there are corrections
in the normalisation of the particles' quantised momentum
\cite{HashTayl}). The action for the bosonic sector of $D=10$ $U(N)$
super Yang Mills reduced to 6-dimensions is (we take $g_{YM}=1$)
\eqn
S=-\frac{1}{4}\int d^6x\ {\rm tr}\left\{(G_{\a\b})^2+
2(\p_{\a}\phi_m+i[B_{\a},\phi_m])^2-[\phi_m,\phi_n]^2\right\}\ ,
\label{4.1}
\eeqn
where
\eqn
G_{\a\b}=\p_{\a}B_{\b}-\p_{\b}B_{\a}+i[B_{\a},B_{\b}]\ ,
\label{4.2}
\eeqn
with $\a,\b=0,...,5$ and $m,n=1,...,4$. The fields
$B_{\a}$ and $\phi_m$ are in the adjoint representation of $U(N)$,
i.e. they take values in the $U(N)$ Lie algebra which we take to be
hermitian. The action (\ref{4.1}) is invariant under the local gauge
transformations
\eqn
\arr{l}
B_{\a}\rightarrow [\O]B_{\a}
\equiv \O B_{\a}\O^{-1}-i\O\p_{\a}\O^{-1}\ ,\\\\
\phi_m\rightarrow [\O]\phi_m
\equiv \O\phi_m\O^{-1}\ ,
\earr
\label{4.3}
\eeqn
where $\O\ \in\ U(N)$.

Our D-brane system is composed by $n$ D-5-branes, each one with a
constant self-dual field strength on the 4-torus in the
$x^{\hat{\a}}$-directions $(\hat{\a}=2,...,5)$ and wrapped in the
$x^2$- and $x^4$-directions with winding numbers $p_i$ and
$\bar{p}_i$, respectively. In order to construct a
$U(N)$ bundle over this $T^4$, the gauge potential $B_{\a}$ and the
scalar fields $\phi_m$ have to satisfy the boundary
conditions\footnote{Note that $B_{\a}(x^{\hb}+L_{\hb})$ is short for
  $B_{\a}(x^0,x^1,...,x^{\hb}+L_{\hb},...,x^4)$ and similarly for the
  other fields.} \cite{tHoo,tHoo1}
\eqn
\arr{l}
B_{\a}(x^{\hb}+L_{\hb})=[\O_{\hb}] B_{\a}(x^{\hb})\equiv
\O_{\hb}B_{\a}(x^{\hb})\O_{\hb}^{-1}
-i\O_{\hb}\p_{\a}\O_{\hb}^{-1}\ ,\\\\
\phi_m(x^{\hb}+L_{\hb})=[\O_{\hb}] \phi_m(x^{\hb})\equiv
\O_{\hb}\phi_m(x^{\hb})\O_{\hb}^{-1}\ ,
\earr
\label{4.4}
\eeqn
where $L_{\ha}=2\pi R_{\ha}$.
Physically this means that the fields are periodic over $T^4$ up to
gauge transformations. The $\O_{\ha}$'s are called multiple transition
functions. Consistency of the boundary conditions (\ref{4.4}) requires
that the $\O_{\ha}$'s obey the periodicity conditions
\eqn
\O_{\ha}(x^{\hb}+L_{\hb}) \O_{\hb}(x) \O^{-1}_{\ha}(x) 
\O^{-1}_{\hb}(x^{\ha}+L_{\ha})=1\equiv center\ of\ U(N)\ ,
\label{4.6}
\eeqn
and that $\O_{\ha}(x)$ is independent of $x^{\ha}$.
If we perform a gauge transformation $\O$ the
fields $B_{\a}$ and $\phi_m$ will change according to (\ref{4.3})
while the $\O_{\ha}$'s transformation rules are
\eqn
\O_{\ha}\rightarrow \O(x^{\ha}=L_{\ha})\O_{\ha}\O^{-1}(x^{\ha}=0)\ .
\label{4.5}
\eeqn
The point now is that we can have a non-trivial bundle with the
fields $B_{\a}$ and $\phi_m$ being very simple while the structure of the
solution is absorbed in the $\O_{\ha}$'s \cite{tHoo1}.

Several examples of non-trivial bundles of this type have been
presented in the literature to explain the gauge theory for D-brane
systems carrying flux in the $U(1)$ part of the theory
\cite{GuraRamg,HashTayl,GuraRamg1,OhtaZhou} (see also \cite{PapaTesc}
for a case where space in not compactified). The
general philosophy is to decompose the condition (\ref{4.6}) in its
$U(1)$ and $SU(N)/Z_N$ parts. The $U(1)$ flux will induce a t'Hooft
twist and the old results for $SU(N)/Z_N$ bundles on a torus may be
used \cite{tHoo,tHoo1,Baal,Baal1}. 

Once we have specified the boundary conditions (\ref{4.4}) for the
$U(N)$ theory describing the system of $n$ D-5-branes, we may study
the corresponding spectrum and analyse a supersymmetric branch of the
theory with a maximum number of massless particles. We shall consider
the simplest case $p_i=\bar{p}_i=1$ for all $i$, and leave
the more general case for an appendix. Let us start by writing some
formulae that will be useful in both cases. We will expand the action
(\ref{4.1}) around a given gauge field background $B_{\a}^0$ solving
the classical equations of motion, i.e. we write the gauge potential
and corresponding field strength as
\eqn
\arr{l}
B_{\a}=B_{\a}^0+A_{\a}\ ,\\\\
G_{\a\b}=G^0_{\a\b}+F_{\a\b}\ ,
\earr
\label{4.7}
\eeqn
with
\eqn
\arr{l}
G^0_{\a\b}=\p_{\a}B_{\b}^0-\p_{\b}B_{\a}^0+i[B_{\a}^0,B_{\b}^0]\ ,\\\\
F_{\a\b}=D_{\a}A_{\b}-D_{\b}A_{\a}+i[A_{\a},A_{\b}]\ ,\\
\earr
\label{4.8}
\eeqn
where $D_{\a}\equiv \p_{\a}+i[B_{\a}^0,\ \ ]$. In the present case the
background field strength is constant, diagonal and only the
$x^{\ha}$-components ($\ha=2,...,5$) are non-vanishing. After some
algebra the following action may be obtained \cite{Baal1}
\eqna
S&=& -\frac{1}{4}\int d^6x 
\ {\rm tr}\left\{-2A^{\a}D^2A_{\a}
-4iA^{\hb}[G_{\hb\ha}^0,A^{\ha}]-2\phi^mD^2\phi_m\right.
\nonumber\\\nonumber\\
&&+2i(D_{\a}A_{\b}-D_{\b}A_{\a})[A^{\b},A^{\a}]
+4i\phi^mD_{\a}[\phi_m,A^{\a}]
\label{4.9}\\\nonumber\\
&&\left. -[A_{\a},A_{\b}]^2-2[A_{\a},\phi_m]^2
-[\phi_m,\phi_n]^2\right\}\nonumber
\eeqna
where the field $A_{\a}$ satisfies the background gauge fixing
condition $D_{\a}A^{\a}=0$.

\subsection{$p_i=\bar{p}_i=1$ case}

We have explained in the introduction that the integers $p_i$ and
$\bar{p}_i$ are the winding numbers of the type $i$ D-5-brane in the
$x^2$- and $x^4$-directions, respectively. Hence, for the case
considered in this subsection we have $n$ coincident D-5-branes, each
one singly wrapped on $T^4$. The
gauge group is $U(n)$ and it is broken to $U(1)^n$ by the background
field strength (see eqn. (\ref{1.3}))
\eqna
G^0_{23}&=&\frac{2\pi}{L_2L_3}{\rm diag}\left( q_1,...,q_n\right)\ ,
\nonumber\\\label{4.10}\\
G^0_{45}&=&
\frac{2\pi}{L_4L_5}{\rm diag}\left(\bar{q}_1,...,\bar{q}_n\right)\ .\nonumber
\eeqna
We shall see that some of the particles
coming from this symmetry breaking process turn out to be
massless. They are described by torons on the $T^4$ directions and are
responsible for the black hole entropy. 

Each D-5-brane carries $U(1)$ fluxes $2\pi q_i$ and
$2\pi\bar{q}_i$ in the $x^2x^3$ and $x^4x^5$ 2-torus,
respectively. These will induce a non-trivial twist in the
$SU(n)/Z_n$ part of the theory. Our bundle is defined by the
following multiple transition functions
\eqn
\O_{\ha}=
\exp{\left(-\pi i \sum_in^i_{\ha\hb}\frac{x^{\hb}}{L_{\hb}}T_i\right)}\ ,
\label{4.11}
\eeqn
where $(T_i)_{ab}=\d_{ia}\d_{ib}$ is the $n\times n$
matrix in the $U(n)$ Lie algebra whose only non-vanishing element is
the $i$-th diagonal entry and
\eqn
\arr{rl}
n^i_{\ha\hb}=&\left(
\arr{cccc}
0&q_i&0&0\\
-q_i&0&0&0\\
0&0&0&\bar{q}_i\\
0&0&-\bar{q}_i&0
\earr
\right)
\earr
\label{4.12}
\eeqn
is the twist tensor due to the $i$-th D-5-brane. The periodicity
conditions (\ref{4.6})
hold because $q_i$ and $\bar{q}_i$ are integers.

The boundary conditions for the gauge potential $B_{\a}$ are now seen
to be
\eqna
B_{\ha}(x^{\hb}+L_{\hb})&=&
\O_{\hb}B_{\ha}(x^{\hb})\O_{\hb}^{-1}+
\frac{\pi}{L_{\ha}}\sum_in^i_{\hb\ha}T_i\ ,
\nonumber\\\label{4.13}\\
B_{\s}(x^{\hb}+L_{\hb})&=&
\O_{\hb}B_{\s}(x^{\hb})\O_{\hb}^{-1}\ ,\nonumber
\eeqna
where $\s=0,1$ and $\O_{\hb}$ is given by (\ref{4.11}). These boundary
conditions are solved by the background field
\eqn
B^0_{\ha}=-\pi \sum_in^i_{\ha\hb} \frac{x^{\hb}}{L_{\ha}L_{\hb}}T_i\ ,
\label{4.14}
\eeqn
with field strength
\eqn
G^0_{\ha\hb}=2\pi\sum_i\frac{n^i_{\ha\hb}}{L_{\ha}L_{\hb}}T_i\ ,
\label{4.15}
\eeqn
which is, as expected, the same as in equation (\ref{4.10}). The
fluctuating fields $A_{\a}$ will now obey the same boundary conditions as the
fields $\phi_m$ in (\ref{4.4}).

We have found the background gauge field for our theory, i.e. our
vacuum state, and the boundary conditions on the quantum fluctuations
around this background. The next step is to expand the fields in the
$U(n)$ Lie algebra:
\eqna
A_{\a}&=&\sum_ia^i_{\a}T_i+\sum_{ij}b^{ij}_{\a}e_{ij}\ ,
\nonumber\\\label{4.16}\\
\phi_m&=&\sum_ic^i_mT_i+\sum_{ij}d^{ij}_me_{ij}\ ,\nonumber
\eeqna
where as before $(T_i)_{ab}=\d_{ia}\d_{ib}$ and 
$(e_{ij})_{ab}=\d_{ia}\d_{jb}$. The fields are taken to be hermitian,
therefore $a^i_{\a}$ and $c^i_m$ are real while 
$b^{ij}_{\a}=(b^{ji}_{\a})^*$ and $d^{ij}_m=(d^{ji}_m)^*$. Expanding
the action (\ref{4.9}) in terms of these field components and keeping
only the quadratic terms we obtain \cite{Baal1}
\eqna
S&=&-\frac{1}{2}\int d^6x \left\{
\sum_i\left( a^i_{\a}M_0a^i_{\a}+c^i_mM_0c^i_m\right)
+2\sum_{i<j}\left( (b_{\s}^{ij})^*M_{ij}b_{\s}^{ij}
\right.\right.
\nonumber\\\label{4.17}\\
&&\left.\phantom{\sum_i}\left. +(b_{\ha}^{ij})^*
\left[\d_{\ha\hb}M_{ij}-4\pi J^{ij}_{\ha\hb}\right] b_{\hb}^{ij}
+(d_m^{ij})^*M_{ij}d_m^{ij}\right)+O(3)\right\}\ ,
\nonumber
\eeqna
where $\s=0,1$ and the operators $M_0$ and $M_{ij}$ are given by
\eqna
M_0&=&\left(\frac{1}{i}\p_{\a}\right)^2=
\p_0^2-\p_1^2-\left(\p_{\ha}\right)^2\ ,
\nonumber\\\label{4.18}\\
M_{ij}&=&\p_0^2-\p_1^2
+\left(\frac{1}{i}\p_{\ha}-\pi J^{ij}_{\ha\hb}x^{\hb}\right)^2\ ,
\nonumber
\eeqna
with
\eqna
J^{ij}_{\ha\hb}&=&\frac{n^i_{\ha\hb}-n^j_{\ha\hb}}{L_{\ha}L_{\hb}}
\equiv \frac{n^{ij}_{\ha\hb}}{L_{\ha}L_{\hb}}
\label{4.19}\\\nonumber\\
&=&
\left(
\arr{cccc}
0&\frac{q_i-q_j}{L_2L_3}&0&0\\
-\frac{q_i-q_j}{L_2L_3}&0&0&0\\
0&0&0&\frac{\bar{q}_i-\bar{q}_j}{L_4L_5}\\
0&0&-\frac{\bar{q}_i-\bar{q}_j}{L_4L_5}&0
\earr
\right)\equiv\left(
\arr{cccc}
0&f^{ij}&0&0\\
-f^{ij}&0&0&0\\
0&0&0&f^{ij}\\
0&0&-f^{ij}&0
\earr
\right)
\nonumber
\eeqna
a self-dual tensor on $T^4$. The boundary conditions on the fields
$A_{\a}$ and $\phi_m$ now read
\eqna
a^i_{\a}(x^{\hb}+L_{\hb})&=&a^i_{\a}(x^{\hb})\ ,
\nonumber\\\nonumber\\
b^{ij}_{\a}(x^{\hb}+L_{\hb})&=&
\exp{\left( -\pi i n^{ij}_{\hb\hg}\frac{x^{\hg}}{L_{\hg}}\right)}
b^{ij}_{\a}(x^{\hb})\ ,
\nonumber\\\label{4.20}\\
c^i_m(x^{\hb}+L_{\hb})&=&c^i_m(x^{\hb})\ ,
\nonumber\\\nonumber\\
d^{ij}_m(x^{\hb}+L_{\hb})&=&
\exp{\left( -\pi i n^{ij}_{\hb\hg}\frac{x^{\hg}}{L_{\hg}}\right)}
d^{ij}_m(x^{\hb})\ .\nonumber
\eeqna

Having found the boundary conditions and quadratic operators
for all the field components we can proceed by identifying the classical
configurations associated with each of these fields. In other
words, we will solve the classical field equations neglecting
interaction terms. We shall then choose an appropriate supersymmetric
branch of the theory with a maximum number of massless fluctuations
around the vacuum. To choose this branch of the theory we will consider the
interaction terms in the Lagrangian. In subsection 4.1.1 below we shall
study the spectrum of the $M_{ij}$ operator acting on $T^4$. These
results were presented in \cite{Baal1} and are reviewed here for the
sake of completeness. In subsection 4.1.2 we shall describe the
classical configurations associated with each field and the
corresponding particle excitations. These results will allow us to
choose a supersymmetric branch of the theory and to compute the
entropy of the corresponding excited D-brane system.

\subsubsection{$M_{ij}$ operator on $T^4$}

In order to find the spectrum of the $M_{ij}$ operator on $T^{4}$ for
given $i$ and $j$, we define the complex coordinates
\eqn
z=(z_1,z_2)=\frac{1}{\sqrt{2}}\left( x_2-ix_3,x_4-ix_5\right)\ ,
\label{4.21}
\eeqn
In these coordinates the worldvolume gauge field is written as
\eqn
A_z=(A_{z_1},A_{z_2})=\frac{1}{\sqrt{2}}
\left( A_2+iA_3,A_4+iA_5\right)\ .
\label{4.22}
\eeqn
Without loss of generality we assume that $f^{ij}\ge 0$ (see
eqn. (\ref{4.19})) and define the positive definite hermitian form
\eqn
H^{ij}(z,w)=2\left( z_1f^{ij}\bar{w}_1+z_2f^{ij}\bar{w}_2\right)=
w^{\dagger}h^{ij}z\ ,
\label{4.23}
\eeqn
where $h^{ij}=2{\rm diag}(f^{ij},f^{ij})$. We define the operators
\eqna
a^{ij}=\left( a^{ij}_1,a^{ij}_2\right)\ ,\ \ \ &
a^{ij}_k=
\frac{1}{i}\left(\p_{\bar{z}_k}+\pi f^{ij}z_{k}\right)\ ,
\nonumber\\\label{4.24}\\
(a^{ij})^{\dagger}=
\left( (a^{ij}_1)^{\dagger},(a^{ij}_2)^{\dagger}\right)\ ,\ \ \ &
(a^{ij}_k)^{\dagger}=
\frac{1}{i}\left(\p_{z_k}-\pi f^{ij}\bar{z}_{k}\right)\ ,
\nonumber
\eeqna
where $k=1,2$. The part of the $M_{ij}$ operator in (\ref{4.18})
acting on $T^4$ may now be written as
\eqn
M_{ij}=\left(\frac{1}{i}\p_{\ha}-\pi J^{ij}_{\ha\hb}x^{\hb}\right)^2=
\sum_k\{a_k^{ij},(a_k^{ij})^{\dagger}\}\ ,
\label{4.25}
\eeqn
and we have
\eqn
[a_k^{ij},(a_l^{ij})^{\dagger}]=2\pi f^{ij}\d_{kl}\ .
\label{4.26}
\eeqn
Thus, the operators $a^{ij}$ and $(a^{ij})^{\dagger}$ may be seen as
annihilation and creation operators, respectively. The ground state is
defined by
\eqn
a^{ij}|0\rangle=0\ ,
\label{4.27}
\eeqn
with eigenvalue $\la_0=4\pi f^{ij}$. The excited states are then
\eqn
|m_1m_2\rangle=\frac{\left( (a_1^{ij})^{\dagger}\right)^{m_1}
\left( (a_2^{ij})^{\dagger}\right)^{m_2}}{\sqrt{m_1!m_2!}}|0\rangle\ ,
\label{4.28}
\eeqn
with
\eqn
\la_{m_1m_2}=4\pi f^{ij}\left( 1+m_1+m_2\right)\ .
\label{4.28a}
\eeqn
The ground state wave function 
$\chi_{_0}(z,\bar{z})=\langle z,\bar{z}|0\rangle$
satisfies
\eqn
a^{ij}\chi_{_0}(z,\bar{z})=0\Rightarrow 
\left(\p_{\bar{z}_k}+\pi f^{ij}z_k\right)\chi_{_0}=0\ ,\ \ \ k=1,2\ ,
\label{4.29}
\eeqn
with $\chi_{_0}$ satisfying the boundary conditions specified
below. Writing
\eqn
\chi_{_0}(z,\bar{z})=\exp{\left(-\frac{\pi}{2}H(z,\bar{z})\right)}
f(z,\bar{z})\ ,
\label{4.30}
\eeqn
equation (\ref{4.29}) becomes $f(z,\bar{z})=f(z)$, i.e. $f$ is
holomorphic. The wave functions $\chi_{_{m_1m_2}}(z,\bar{z})$ for the
excited states may be obtained by acting with the creation operators
on the ground state wave function. 

The operators $M_{ij}$ in the
action (\ref{4.17}) act on the fields $b_{\a}^{ij}$ and $d_m^{ij}$,
both satisfying the same boundary conditions. The dependence of these
fields on $T^4$ will be determined by the ground state wave function
$\chi_{_0}(z,\bar{z})$ and the excited states wave
functions $\chi_{_{m_1m_2}}(z,\bar{z})$. Hence, the function
$\chi_{_0}(z,\bar{z})$ obeys the following boundary conditions
\eqn
\chi_{_0}(x^{\hb}+L_{\hb})=
\exp{\left( -\pi i n^{ij}_{\hb\hg}\frac{x^{\hg}}{L_{\hg}}\right)}
\chi_{_0}(x^{\hb})\ .
\label{4.31}
\eeqn
Note that for some function $\chi_{_0}(z,\bar{z})$ obeying these
boundary conditions and using equation (\ref{4.28}) it follows that
the functions $\chi_{_{m_1m_2}}(z,\bar{z})$ also obey the same
boundary conditions. We now define the vector
\eqna
q=(q_1,q_2)\ ,\ \ \ \ 
&q_1=\frac{1}{\sqrt{2}}\left(m_2L_2-im_3L_3\right)\ ,
\label{4.32}\\
&q_2=\frac{1}{\sqrt{2}}\left(m_4L_4-im_5L_5\right)\ ,\nonumber
\eeqna
taking values on the $T^4$ lattice. The condition (\ref{4.31}) reads
\eqn
f(z+q)=\a(q)\exp{\left(\frac{\pi}{2}H(q,q)+\pi H(z,q)\right)}f(z)\ ,
\label{4.33}
\eeqn
where
\eqn
\a(q)=\exp{
\left(\pi i \sum_{\ha<\hb}m_{\ha}n^{ij}_{\ha\hb}m_{\hb}\right)}\ .
\label{4.34}
\eeqn
The holomorphic functions $f$ satisfying (\ref{4.33}) are called in
the mathematical literature $\Theta$-functions of type $(H,\a)$
\cite{Igusa,Hano}. They form a complex linear space $L(H,\a)$ of
dimension
\eqn
|Pf(n^{ij})|=
\frac{1}{8}\e_{\ha\hb\hg\hth}n^{ij}_{\ha\hb}n^{ij}_{\hg\hth}=
|(q_i-q_j)(\bar{q}_i-\bar{q}_j)|=n^{ij}_L\bar{n}^{ij}_L\ .
\label{4.35}
\eeqn
Thus, there are $n^{ij}_L\bar{n}^{ij}_L$ orthogonal ground state wave
functions that we write as
\eqn
\chi_{_0}^r(z,\bar{z})\ ,\ \ \ \ \ r=1,...,n^{ij}_L\bar{n}^{ij}_L\ .
\label{4.36}
\eeqn
This is the origin of the Landau degeneracy in the gauge theory
description of our D-brane system. For a given $i,j$ pair each of
these functions will be associated with a massless toronic excitation of the
D-brane system, giving exactly the entropy formula (\ref{2.7}).

\subsubsection{Particle spectrum}
 
In this subsection we shall describe the particle excitations associated
with each field in the expansion (\ref{4.16}). In other words, we will
consider the free Lagrangian in
(\ref{4.17}). Consider first the gauge fields $a^i_{\a}$ associated
with the remaining $U(1)^n$ gauge freedom. The classical equations of
motion are solved by the zero modes of the operator $M_0$ in
(\ref{4.17})
\eqn
a^i_{\a}=\sum_p e_{\a}(p)\exp{\left(ip_{\b}x^{\b}\right)}\ ,\ \ \ 
p^2=0\ .
\label{4.37}
\eeqn
The gauge condition $D_{\a}A^{\a}=0$ becomes for these field
components the usual Lorentz gauge condition $\p_{\a}a^i_{\a}=0$. We can
further impose the Coulomb gauge condition $A_0=0$. Thus, the fields
$a^i_{\a}$ describe massless vector gauge particles in 6 dimensions
corresponding to $4n$ degrees of freedom. The analysis for the $c^i_m$
scalar fields is similar. These fields describe $4n$ massless scalar
particles. 

We now turn to the fields corresponding to the non-diagonal elements
of the Lie algebra. Consider first the $d_m^{ij}$ components of the
scalar fields $\phi_m$. The classical equations for these fields are
solved by
\eqn
d_m^{ij}=\sum_{rm_1m_2}\chi^r_{_{m_1m_2}} 
\left( {\cal D}_m^{ij}\right)^r_{m_1m_2}(x^{\s})\ ,
\label{4.38}
\eeqn
where $\s=0,1$ and 
$\left( {\cal D}_m^{ij}\right)^r_{m_1m_2}(x^{\s})$ is a complex
conformal field satisfying the equation of motion
\eqn
\left( \Box_{\s}-\la_{m_1m_2}\right)
\left( {\cal D}_m^{ij}\right)^r_{m_1m_2}(x^{\s})=0\ ,
\label{4.39}
\eeqn
i.e.
\eqn
\left( {\cal D}_m^{ij}\right)^r_{m_1m_2}=
\sum_{p_{\s}}\left( e_m^{ij}\right)^r_{m_1m_2}(p_{\s})
\exp{\left(ip_{\s}x^{\s}\right)}\ ,\ \ \ p_{\s}^2=-\la_{m_1m_2}\ .
\label{4.40}
\eeqn
Thus, for each pair $m_1,m_2$ we have $8n^{ij}_L\bar{n}^{ij}_L$ massive
particles with mass given by $m^2=\la_{m_1m_2}$ and described
by a two dimensional field theory. These fields correctly reproduce
the string excitations (\ref{3.12}) if we make the
identification \cite{HashTayl}
\eqn
\frac{\e}{\a'}=\frac{\th_i-\th_j}{\pi\a'}\equiv
\frac{\tan{\th_i}-\tan{\th_j}}{\pi\a'}=4\pi f^{ij}\ .
\label{4.41}
\eeqn
The role that the Born-Infeld description of the D-brane dynamics may
have in solving this discrepancy was discussed in
\cite{HashTayl}. 

We proceed by considering the $b_{\a}^{ij}$ components of the gauge
field $A_{\a}$. Start by writing the quadratic operator acting
on the $b_{\ha}^{ij}$ fields in terms of the $z$ coordinates defined
in (4.21-22) \cite{Baal1}
\eqn
S\sim  -\int d^6x
\sum_{i<j}\left\{
(b_{z_k}^{ij})^*\left[ M_{ij}-4\pi f^{ij}\right] b_{z_k}^{ij}
+(b_{\bar{z}_k}^{ij})^*\left[ M_{ij}+4\pi f^{ij}\right]
b_{\bar{z}_k}^{ij} \right\}\ .
\label{4.42}
\eeqn
Hence, the spectrum of the operator 
$\left(\d_{\hb\ha}M_{ij}-4\pi i J^{ij}_{\hb\ha}\right)$ in
(\ref{4.17}) is
\eqna
\la^-_{m_1m_2}=\la_{m_1m_2}-4\pi f^{ij}\ ,\ \ \ \ 
&b_{z_k}^{ij}=\chi^r_{m_1m_2}\ \ \ (k=1,2)\ ,
\nonumber\\\label{4.43}\\
\la^+_{m_1m_2}=\la_{m_1m_2}+4\pi f^{ij}\ ,\ \ \ \ 
&b_{\bar{z}_k}^{ij}=\chi^r_{m_1m_2}\ \ \ (k=1,2)\ .\nonumber
\eeqna
For simplicity we consider the cases
$b_{\bar{z}_k}^{ij}=0$ or $b_{z_k}^{ij}=0$ separately. In the former
case the classical equations of motion for the $b_{z_k}^{ij}$ fields
are solved by
\eqn
b_{z_k}^{ij}=\sum_{rm_1m_2}\chi^r_{_{m_1m_2}} 
\left( {\cal B}_{z_k}^{ij}\right)^r_{m_1m_2}(x^{\s})\ ,
\ \ \ (k=1,2),
\label{4.44}
\eeqn
with
\eqn
\left( {\cal B}_{z_k}^{ij}\right)^r_{m_1m_2}=
\sum_{p_{\s}}\left( e_{z_k}^{ij}\right)^r_{m_1m_2}(p_{\s})
\exp{\left(ip_{\s}x^{\s}\right)}\ ,\ \ \ p_{\s}^2=-\la^-_{m_1m_2}\ .
\label{4.45}
\eeqn
In the gauge $A_0=0$, the background gauge condition $D_iA_i=0$
becomes
\eqn
\p_1b^{ij}_1+ia^{ij}_kb^{ij}_{z_k}=0\ .
\label{4.46}
\eeqn
The fields $b_1^{ij}$ obey the same equations of motions as the scalar
fields $d_m^{ij}$. Hence
\eqn
b_1^{ij}=\sum_{rm_1m_2}\chi^r_{_{m_1m_2}} 
\left( {\cal B}_1^{ij}\right)^r_{m_1m_2}(x^{\s})\ ,
\label{4.47}
\eeqn
with
\eqn
\left( {\cal B}_1^{ij}\right)^r_{m_1m_2}=
\sum_{p_{\s}}\left( e_1^{ij}\right)^r_{m_1m_2}(p_{\s})
\exp{\left(ip_{\s}x^{\s}\right)}\ ,\ \ \ p_{\s}^2=-\la_{m_1m_2}\ .
\label{4.48}
\eeqn
The gauge condition (\ref{4.46}) becomes
\eqn
p_1\left( e_1^{ij}\right)^r_{m_1m_2}
+2\pi f^{ij}\left[
\sqrt{m_1+1}\left( e_{z_1}^{ij}\right)^r_{m_1+1,m_2}+
\sqrt{m_2+1}\left( e_{z_2}^{ij}\right)^r_{m_1,m_2+1}\right] =0\ ,
\label{4.49}
\eeqn
completely determining the fields $b_1^{ij}$. Note
that since $\la_{m_1m_2}=\la^-_{m_1+1,m_2}=\la^-_{m_1,m_2+1}$ a given
excitation $b_{z_k}^{ij}$ induces a $b_1^{ij}$ field with the
correct mass. Thus, the complex conformal fields 
$\left( {\cal B}_{z_k}^{ij}\right)^r_{m_1m_2}$ describe for each pair
$m_1m_2$, $4n^{ij}_L\bar{n}^{ij}_L$ particles with mass given by
$m^2=\la^-_{m_1m_2}$. With the identification (\ref{4.41}) we
correctly reproduce the spectrum associated with the string states
(\ref{3.10}). Note that if all excitations are in their vacuum
state but the $m_1=m_2=0$ torons then $b_1^{ij}=0$ and the field
configuration (\ref{4.44}) describes $4n^{ij}_L\bar{n}^{ij}_L$
massless torons. These are the particle-like excitations responsible
for the bosonic contribution to the entropy formula as they will allow
us to define a supersymmetric branch of the theory.

In the case $b_{z_k}^{ij}=0$, the complex conformal fields 
$\left( {\cal B}_{\bar{z}_k}^{ij}\right)^r_{m_1m_2}(x^{\s})$
associated with the field $b_{\bar{z}_k}^{ij}$ are given by
\eqn
\left( {\cal B}_{\bar{z}_k}^{ij}\right)^r_{m_1m_2}=
\sum_{p_{\s}}\left( e_{\bar{z}_k}^{ij}\right)^r_{m_1m_2}(p_{\s})
\exp{\left(ip_{\s}x^{\s}\right)}\ ,\ \ \ p_{\s}^2=-\la^+_{m_1m_2}\ .
\label{4.50}
\eeqn
The gauge condition $D_iA_i=0$ becomes
\eqn
\p_1b^{ij}_1+i\left(a^{ij}_k\right)^{\dagger}b^{ij}_{\bar{z}_k}=0\ ,
\label{4.51}
\eeqn
and the fields in (\ref{4.48}) are determined by
\eqn
p_1\left( e_1^{ij}\right)^r_{m_1m_2}+
\sqrt{m_1}\left( e_{\bar{z}_1}^{ij}\right)^r_{m_1-1,m_2}+
\sqrt{m_2}\left( e_{\bar{z}_2}^{ij}\right)^r_{m_1,m_2-1}=0\ ,
\label{4.52}
\eeqn
These fields correctly reproduce the string states (\ref{3.11}).

\subsubsection{Supersymmetric branch}

Now that we have found the different particle excitations associated
with the fields leaving on the D-branes we can choose a supersymmetric
configuration describing the excited D-brane system. We will consider
the case when we have $N_i$ D-5-branes of a given type $i$ and comment
later on the case when we have a single brane of such type wrapped in the
$x^1$-direction with winding number $N_i$. We have to realize that the
fields $a^i_{\a}$ and $c^i_m$ are in the adjoint representation of
$U(N_i)$ and the fields $b^{ij}_{\a}$ and $d^{ij}_m$ in the
fundamental representation of $U(N_i)\times U(\bar{N}_j)$. In other
words, the labels $i$ and $ij$ now carry some group theory indices. In
order to find a supersymmetric branch of the theory with a maximum
number of massless particles consider first the commutator term
$[A_{\a},\phi_m]^2$ in the action (\ref{4.9}). The fields
$c^i_m$ contribute with the following mass term to the $b^{ij}_{\a}$
fields (we are being schematic here and dropping $U(N_i)$
group theory indices)

\eqn
S\sim 
-\int d^6x\sum_{i<j}\left( b_{\a}^{ij}\right)^{\dagger}
\left(c^i-c^j\right)^2 b_{\a}^{ij}\ .
\label{4.53}
\eeqn
Physically this means that if the scalars $c^i_m$ are excited the
D-5-branes will oscillate in the transverse space directions, giving a
mass term to the strings stretching between different branes and
therefore to the $b_{\a}^{ij}$ fields. Inversely, if there are
massless toronic excitations due to the $b_{\a}^{ij}$ fields, there
will be mass terms for the massless scalars and the branes become
bounded, i.e. we have a bound state \cite{Mald}. Our excited D-brane
system is described by the latter picture.  

Next consider the scalar fields
$d_m^{ij}$. Since these are massive fields they are not relevant for
the entropy counting. They  break all the
supersymmetry. Thus, all the components of the scalar fields $\phi_m$
are in their vacuum state. The remaining massless excitations
correspond to the massless gauge bosons $a^i_{\a}$ and the massless
torons $b_{\a}^{ij}$.

Having ruled out from our counting the scalar fields $\phi_m$, we
have to impose supersymmetry of our field configuration. This means
that not all the degrees of freedom in the $a^i_{\a}$ and
$b_{\a}^{ij}$ fields will be independent. We start by imposing the
conditions for our fields to carry left-moving momentum in the
$x^1$-direction
\eqna
&&b_{z_k}^{ij}=
\sum_r\chi_{_0}^r\left( {\cal B}_{z_k}^{ij}\right)^r_0(x^-)\ ,\ \ \ 
b_{\bar{z}_k}^{ij}=b_0^{ij}=b^{ij}_1=0\ ,
\nonumber\\\label{4.54}\\
&&a_{\ha}^i=a_{\ha}^i(x^{\hb},x^-)\ ,\ \ \ a^i_0=a^i_1=0\ ,\nonumber
\eeqna
where $x^-=\frac{1}{\sqrt{2}}(x^0-x^1)$. Note that we are working in
the gauge $A_0=0$, and further imposed the
condition $a^i_1=0$.\footnote{Alternatively we may write 
$a_{\a}^i=a_{\a}^i(x^-)$, i.e. we neglect the massive Ka\l u\.{z}a-Klein
modes arising from the compactification of our theory to $1+1$
dimensions, and the condition $a_1^i=0$ follows from the Lorentz
gauge condition $\p_{\a}a_{\a}^i=0$ in the Coulomb gauge
$a_0^i=0$.}  With the fields as in (\ref{4.54}) we have
(recall that $G_{\a\b}=G_{\a\b}^0+F_{\a\b}$)
\eqn
G_{01}=F_{01}=0\ ,\ \ \ G_{0\ha}=F_{0\ha}=-F_{1\ha}=-G_{1\ha}\ .
\label{4.55}
\eeqn
The supersymmetry variation of the gaugino field (in $10D$ spinor
language) is given by  
\eqn
\d\Psi=-\frac{1}{4}G_{\a\b}\Gamma^{\a\b}\e 
-\frac{1}{2}\left( \p_{\a}\phi_m+i[B_{\a},\phi_m]\right)\Gamma^{\a m}\e
-\frac{i}{4}[\phi_m,\phi_n]\Gamma^{mn}\e\ ,
\label{4.56}
\eeqn
where the last two terms vanishe for our field configuration. The
contribution of the field strength components in (\ref{4.55}) to
$\d\Psi$ vanishes if $(\Gamma^0-\Gamma^1)\e=0$. This
condition breaks $1/2$ of
the worldvolume supersymmetry. To preserve $1/4$ of the
supersymmetry, note that the background field strength is self-dual on
$T^4$. Thus, by requiring self-duality of the field strength
$F_{\ha\hb}$ associated with the fluctuating fields, we can cancel the
contribution of $G_{\ha\hb}$ to $\d\Psi$. This follows by imposing the
condition
$\Gamma^{2345}\e=\e$ on the spinor $\e$. We remark that in order to
vanish the contribution of $G^0_{\ha\hb}$ to $\d\Psi$ we need this
condition on the spinor $\e$ because $G^0_{\ha\hb}$ is not in the $U(1)$
center of the group and therefore the non-linear realization of the
worldvolume supersymmetry transformations can not be used
\cite{HarvMoore}. Thus, we are bound to conclude that a different
condition on $F_{\ha\hb}$ would most likely break further
supersymmetry which is not desirable. 

We were not able to solve the self-duality condition on $F_{\ha\hb}$
exactly. However, we solved this condition to second order in the
fields. Starting with the free theory, i.e. neglecting interactions,
the field configuration describing our excited D-brane system is
\eqna
&&b_{z_k}^{ij}=
\sum_r\chi_{_0}^r\left( {\cal B}_{z_k}^{ij}\right)^r_0(x^-)\ ,\ \ \ 
b_{\bar{z}_k}^{ij}=b_0^{ij}=b^{ij}_1=0\ ,
\nonumber\\\label{4.57}\\
&&a_{\ha}^{i}=0\ ,\ \ \ a_0^{i}=a_1^{i}=0\ .\nonumber
\eeqna
Thus the fields $a^{i}_{\a}$ are in their vacuum state. The field
strength $F_{\a\b}$ may be written as
\eqn
F_{\a\b}=D_{\a}A_{\b}-D_{\b}A_{\a}+i[A_{\a},A_{\b}]
\equiv F_{\a\b}^{(F)}+i[A_{\a},A_{\b}]\ .
\label{4.58}
\eeqn
Equation (\ref{4.29}) may be used to show that the free toronic
excitations are self-dual on $T^4$, i.e. 
$F^{(F)}_{\ha\hb}=\frac{1}{2}\e_{\ha\hb\hg\hth}F^{(F)}_{\hg\hth}$.
Note that
we could have $a_{\ha}^{i}=a_{\ha}^{i}(x^-)$ and the free
theory would still be self-dual. We will comment on this later. In
order to have the self-duality condition for the full $F_{\ha\hb}$
field, i.e. considering the commutator term in (\ref{4.58}), we impose
the condition
\eqn
A_{\hb}^{(2)}=\left( D^2\right)^{-1}D_{\ha}T_{\ha\hb}\ ,
\label{4.59}
\eeqn
where $T_{\ha\hb}$ is quadratic in the free fields in (\ref{4.57}) and
it is an antiself-dual tensor on $T^4$. In components
\eqn
T_{\ha\hb}^{ij}=-i\sum_{k\ne i,j}\left[\left(
b_{\ha}^{ik}b_{\hb}^{kj}-b_{\hb}^{ik}b_{\ha}^{kj}\right)
-\frac{1}{2}\e_{\ha\hb\hg\hth}\left( 
b_{\hg}^{ik}b_{\hth}^{kj}-b_{\hth}^{ik}b_{\hg}^{kj}\right)\right]\ .
\label{4.60}
\eeqn
Notice that the $i,j,k$ labels contain group theory indices. The
$T_{\ha\hb}^{i_aj_{\bar{b}}}$ components are in the fundamental
representation of $U(N_i)\times U(\bar{N}_j)$ and give corrections to the
fields $b_{\ha}^{i_aj_{\bar{b}}}$. The 
$T_{\ha\hb}^{i_ai_{\bar{b}}}\equiv T_{\ha\hb}^{i_{ab}}$ components are
in the adjoint representation of $U(N_i)$ and give corrections to the
fields $a_{\ha}^{i_{ab}}$. We may proceed iteratively to guarantee
self-duality to any order in
the free fields in (\ref{4.57}). It is probably not surprising that we
were not able to show exactly that our field configuration preserves
some supersymmetry. The problem is that the super Yang Mills action
does not reproduce the correct interactions between the fields on the
D-branes. If we had a well defined non-Abelian Born-Infeld action, it
could be that the corresponding torons would be modified (this is in
fact expected in order to have a perfect agreement with the string
theory spectrum \cite{HashTayl}). Also, both actions above
are an approximation to slow varying fields, i.e. the derivatives of
the field strength are neglected. These facts may be preventing an
exact solution of the problem.

In conclusion, the field configuration (\ref{4.57}) is described
quantum mechanically by the left moving sector of a conformal field
theory with $4\sum_{i<j}N_iN_jn_L^{ij}\bar{n}_L^{ij}$ massless
bosonic particles. Supersymmetry requires that we have the same number
of fermionic particles. Thus, we have correctly reproduced the number
of degrees of
freedom required to explain the entropy formula. Two final remarks are
in order: Firstly, as noted in the previous paragraph we could have
excited the $a^i_{\ha}$ fields in the free theory. In the interacting
theory we can still excite the diagonal elements of these fields (more
generally a maximal set of $N_i$ commuting fields). The self-duality
condition on the field strength could still be satisfied by an
appropriate redefinition of $T_{\ha\hb}^{ij}$ in
(\ref{4.60}). However, we could not excite more than these $4N_i$
particles. The corresponding contribution to the entropy is subleading
and the result remains the same ($N_iN_j>>N_k$). Secondly, the same
result as in the string theory description follows if we have a single
D-5-brane of each type wrapped 
$N_i$ times along the $x^1$-direction. In this case we would have
$4\sum_{i<j}n_L^{ij}\bar{n}_L^{ij}$ bosonic and fermionic torons with
momentum quantised in units of $(N_iN_jR_1)^{-1}$ for each $i,j$
pair.

\section{Conclusion}

In this paper we have studied a 5-dimensional black hole associated
with an excited D-brane system described be several coincident
D-5-branes with a self-dual field strength on a 4-torus and carrying
left-moving momentum along a common string. We have analysed the
spectrum of this system either from the string theory perspective and
the gauge theory perspective. Both descriptions give the same answer
for the excited D-brane system entropy matching the Bekenstein-Hawking
entropy formula. These results arise from rather different mathematical
approaches. In the string theory picture the entropy area law is
reproduced by  counting the Landau levels for strings stretching
between different branes, while in the gauge theory picture by
counting the number of massless torons. In the latter description the
number of torons arises from highly non-trivial results on
$\Theta$-functions on $T^4$. As such, our work provides another
example of the fascinating interplay between string theory and gauge
theory. Also, it is quite interesting that old results which
constituted an attempt to understand quark confinement and
non-perturbative aspects of QCD, are now placed in the context of
black hole physics.

There are several unresolved problems that should fit in the general
scheme of things. Firstly, once the non-abelian Born-Infeld action is
known one may expect to find the corresponding BI-torons. These should
yield a perfect agreement with the string theory spectrum
\cite{HashTayl}. It may also prove to be useful to show exactly the
supersymmetry of our system. Secondly, although we have not solved the
number theory problem (\ref{3.18}) we expect it to reproduce the correct
entropy formula. Finally, another problem is to generalise the results
of these paper to the near-extreme black hole case. This has been done
in \cite{CostaPerry} for a different case.

\section*{Acknowledgements}

One of us (M.S.C.) acknowledges the financial support of
JNICT (Portugal) under programme PRAXIS XXI.

\newpage
\section*{Appendix: $p_i\ne 1\ne \bar{p}_i$ case}
\news
\renewcommand{\theequation}{A.\arabic{equation}}

In this appendix we shall describe the modifications required to
treat the general case when the winding numbers of the type $i$
D-5-branes are different from unity. Consider first the case $p_i\ne 1$
but $\bar{p}_i=1$. A D-brane with winding number $p_i$ in a compact
direction is described by a $U(p_i)$ gauge theory broken to
$U(1)^{p_i}$ by Wilson lines or non-trivial boundary conditions
satisfied by the fields in the compact direction
\cite{Polc1,Hash}. Since we have $n$ D-5-branes on top of each other we
start with $U\left( \sum_ip_i\right)$ gauge group broken to
$\prod_i\otimes U(p_i)$ by the background gauge field. Each
D-5-brane wraps around the $x^2$-direction $p_i$ times, breaking each 
$U(p_i)$ factor to $U(1)^{p_i}$. The general case $\bar{p}_i\ne 1$
may be treated similarly. We would have to consider the group
$U(p_i\bar{p}_i)$ for each D-5-brane. Since this case does not bring
any new feature we will describe below the simpler case $\bar{p}_i=1$.

As in subsection 4.1 we start by writing the background field strength
\eqna
G^0_{23}&=&\frac{2\pi}{L_2L_3}
\left(\underbrace{\frac{q_1}{p_1},...,
\frac{q_1}{p_1}},...,
\underbrace{\frac{q_n}{p_n},...,\frac{q_n}{p_n}}\right)\ ,
\nonumber\\
&&{\footnotesize \ \ \ \ \ \ \ \ \ \ \ \ p_1\ times\ \ \ \ \ \ \ p_n\ times}
\label{A.1}\\
G^0_{45}&=&\frac{2\pi}{L_4L_5}
\left( \overbrace{\ \bar{q}_1,...,\bar{q}_1\ },...,
\overbrace{\ \bar{q}_n,...,\bar{q}_n\ }\right)\ .\nonumber
\eeqna
Each D-5-brane carries a flux $2\pi q_i$ and $2\pi p_i\bar{q}_i$ in
the $x^2x^3$ and $x^4x^5$ 2-torus, respectively. The multiple
transition functions for a $U(\sum_ip_i)$ bundle of this type are
\eqn
\O_{\ha}={\rm Diag}\left(\O_{\ha}^{(1)},..., \O_{\ha}^{(n)}\right)\ ,
\label{A.2}
\eeqn
where the $\O_{\ha}^{(i)}$ are the $p_i\times p_i$ matrices
\cite{HashTayl}
\eqn
\arr{ll}
\O_2^{(i)}=\O'^{(i)}_2V_i\ , \ \ \ \ &\O_4^{(i)}=\O'^{(i)}_4\ ,\\\\
\O_3^{(i)}=\O'^{(i)}_3U_i^{q_i}\ , &\O_5^{(i)}=\O'^{(i)}_5\ ,
\earr
\label{A.3}
\eeqn
with
\eqna
U_i&=& diag\left( 1, e^{2\pi i\frac{1}{p_i}},...,
e^{2\pi i\frac{p_i-1}{p_i}}\right)\ ,\nonumber\\\nonumber\\
V_i&=& \left(
\arr{ccccc}
0&1&^.&&\\
^.&0&1&^.&\\
&^.&^.&^.&^.\\
^.&^.&&^.&1\\
1&^.&&^.&0
\earr
\right)\ ,\label{A.4}\\\nonumber\\
\O'^{(i)}_{\ha}&=&
\exp{\left( -\pi i n^{i}_{\ha\hb}\frac{x^{\hb}}{L_{\hb}}{\bf 1}_i\right)}\ .
\nonumber
\eeqna
The twist tensor $n^i_{\ha\hb}$ is given by
\eqn
n^i_{\ha\hb}=\left(
\arr{cccc}
0&\frac{q_i}{p_i}&0&0\\
-\frac{q_i}{p_i}&0&0&0\\
0&0&0&\bar{q}_i\\
0&0&-\bar{q}_i&0
\earr
\right)\ ,
\label{A.5}
\eeqn
and ${\bf 1}_i$ is the $i$-dimensional unit matrix. The consistency
conditions (\ref{4.6}) for the $\O_{\ha}$'s may be seen to hold. The
boundary conditions (\ref{4.4}) are now solved by the background field
\eqn
B_{\ha}^0={\rm Diag}\left( B_{\ha}^{0(1)},...,B_{\ha}^{0(n)}\right)\ ,
\label{A.6}
\eeqn
with
\eqn
B_{\ha}^{0(i)}=-\pi n^i_{\ha\hb}\frac{x^{\hb}}{L_{\hb}L_{\ha}}{\bf 1}_i\ .
\label{A.7}
\eeqn
The corresponding field strength is given by (\ref{A.1}).

We proceed by expanding the fluctuating fields in the $U(\sum_ip_i)$
Lie algebra. As before we write
\eqna
A_{\a}&=&\sum_ia^i_{\a}T_i+\sum_{ij}b^{ij}_{\a}e_{ij}\ ,
\nonumber\\\label{A.8}\\
\phi_m&=&\sum_ic^i_mT_i+\sum_{ij}d^{ij}_me_{ij}\ ,
\nonumber
\eeqna
but now $T_i$ and $e_{ij}$ are $p_i\times p_i$ and $p_i\times p_j$
matrices, respectively. Thus, the fields $a^i_{\a}$ and $c^i_m$ are in
the adjoint representation of $U(p_i)$ and the fields $b_{\a}^{ij}$
and $d_m^{ij}$ $(i<j)$ in the 
fundamental representation of $U(p_i)\otimes U(\bar{p}_j)$ (note that
$b_{\a}^{ij}=(b_{\a}^{ji})^{\dagger}$ and
$d_m^{ij}=(d_m^{ji})^{\dagger}$) \footnote{There is a possible confusion
with the notation here. We are considering the case $\bar{p}_i=1$ for 
all $i$. Thus, we mean by $U(\bar{p}_j)$ the anti-fundamental 
representation of $U(p_j)$.}. With the observation that the
$i$ and $j$ labels contain group theory indices, the result of
expanding the action in terms of the field components is the same as
in (\ref{4.17}). The quadratic operators acting on the fields are also
as in (\ref{4.18}), but the self-dual tensor $J_{\ha\hb}^{ij}$ is now
given by
\eqna
J^{ij}_{\ha\hb}&=&
\frac{n^i_{\ha\hb}-n^j_{\ha\hb}}{L_{\ha}L_{\hb}}\equiv
\left(
\arr{cccc}
0&\frac{n^{ij}_{23}}{p_ip_jL_2L_3}&0&0\\
\frac{n^{ij}_{32}}{p_ip_jL_2L_3}&0&0&0\\
0&0&0&\frac{n^{ij}_{45}}{L_4L_5}\\
0&0&\frac{n^{ij}_{54}}{L_4L_5}&0
\earr
\right)\label{A.9}\\\nonumber\\
&=&\left(
\arr{cccc}
0&\frac{p_jq_i-p_iq_j}{p_ip_jL_2L_3}&0&0\\
-\frac{p_jq_i-p_iq_j}{p_ip_jL_2L_3}&0&0&0\\
0&0&0&\frac{\bar{q}_i-\bar{q}_j}{L_4L_5}\\
0&0&-\frac{\bar{q}_i-\bar{q}_j}{L_4L_5}&0
\earr
\right)\equiv\left(
\arr{cccc}
0&f_{ij}&0&0\\
-f_{ij}&0&0&0\\
0&0&0&f_{ij}\\
0&0&-f_{ij}&0
\earr
\right)\ .\nonumber
\eeqna
The boundary conditions on the $a^i_{\a}$ fields are
\eqna
&&(a_{\a}^i)_{ab}(x^2+L_2)=(a^i_{\a})_{a+1,b+1}(x^2)\ ,
\nonumber\\\nonumber\\
&&(a_{\a}^i)_{ab}(x^3+L_3)=
e^{2\pi i q_i\frac{a-b}{p_i}}(a^i_{\a})_{ab}(x^3)\ ,
\label{A.10}\\\nonumber\\
&&(a_{\a}^i)_{ab}(x^{\hb}+L_{\hb})=(a^i_{\a})_{ab}(x^{\hb})\ ,\ \ \ 
\hb=4,5\ ,\nonumber
\eeqna
where $a,b=1,...,p_i$. The fields $c^i_m$ obey similar boundary
conditions. The fields $b_{\a}^{ij}$ satisfy the following boundary
conditions 
\eqna
&&(b_{\a}^{ij})_{a\bar{b}}(x^2+L_2)=
\exp{\left(-\pi in^{ij}_{23}\frac{x^3}{p_ip_jL_3}\right)}
(b^{ij}_{\a})_{a+1,\overline{b+1}}(x^2)\ ,
\nonumber\\\nonumber\\
&&(b_{\a}^{ij})_{a\bar{b}}(x^3+L_3)=
\exp{\left(-\pi in^{ij}_{32}\frac{x^2+2A_{a\bar{b}}}{p_ip_jL_2}\right)}
(b^{ij}_{\a})_{a\bar{b}}(x^3)\ ,\label{A.11}\\\nonumber\\
&&(b_{\a}^{ij})_{a\bar{b}}(x^{\hb}+L_{\hb})=
\exp{\left(-\pi in^{ij}_{\hb\hg}\frac{x^{\hg}}{L_{\hg}}\right)}
(b^{ij}_{\a})_{a\bar{b}}(x^{\hb})\ ,\ \ \ 
\hb=4,5\ ,\nonumber
\eeqna
where
\eqn
A_{a\bar{b}}=L_2\frac{p_jq_i(a-1)-p_iq_j(\bar{b}-1)}{p_jq_i-p_iq_j}\ ,
\label{A.12}
\eeqn
$a=1,...,p_i$ and $\bar{b}=\bar{1},...,\bar{p}_i$.
The fields $d^{ij}_m$ obey similar boundary conditions.

We proceed by studying the spectrum of the $M_{ij}$ operator on $T^4$
subjected to the boundary conditions (\ref{A.11}) and the massless
particle spectrum of our gauge theory.

\subsubsection*{$M_{ij}$ operator on $T^4$}

The crucial difference between this case and the one presented in the
subsection 4.1, is that the boundary conditions (\ref{A.11}) relate
different field components. The analysis of the spectrum of the
$M_{ij}$ operator in terms of the creation and annihilation operators
goes through as before. The only difference is that the ground state
wave function is defined by
\eqn
(\chi_{_0})_{a\bar{b}}(z,\bar{z})=
\exp{\left( -\frac{\pi}{4}H^{ij}(z-\bar{z},z_0+\bar{z}_0)
-\frac{\pi}{2}H^{ij}(z+z_0,z+z_0)\right)}g(z)\ ,
\label{A.13}
\eeqn
where $z_0=\frac{1}{\sqrt{2}}(A_{a\bar{b}},0)$ and the $a,\bar{b}$ indices are
appropriate for a given field component $(b_{\a}^{ij})_{a\bar{b}}$ with
$A_{a\bar{b}}$ given as in (\ref{A.12}). For such a field component 
the boundary conditions on the function 
$(\chi_{_0})_{a\bar{b}}(z,\bar{z})$ (that describes the dependence on $T^4$)
are
\eqna
&&(\chi_{_0})_{a\bar{b}}(x^2+p_ip_jL_2)=
\exp{\left(-\pi in^{ij}_{23}\frac{x^3}{L_3}\right)}
(\chi_{_0})_{a\bar{b}}(x^2)\ ,\nonumber\\\nonumber\\
&&(\chi_{_0})_{a\bar{b}}(x^3+L_3)=
\exp{\left(-\pi in^{ij}_{32}\frac{x^2+2A_{a\bar{b}}}{p_ip_jL_2}\right)}
(\chi_{_0})_{a\bar{b}}(x^3)\ ,\label{A.14}\\\nonumber\\
&&(\chi_{_0})_{a\bar{b}}(x^{\hb}+L_{\hb})=
\exp{\left(-\pi in^{ij}_{\hb\hg}\frac{x^{\hg}}{L_{\hg}}\right)}
(\chi_{_0})_{a\bar{b}}(x^{\hb})\ ,\ \ \ 
\hb=4,5\ .\nonumber
\eeqna
Thus, apart from the $A_{a\bar{b}}$ shift in the $x^2$-direction the problem
is similar to the previous case. We just have to realize that we have
a periodicity $p_ip_jL_2$ along the $x^2$-direction instead of
$L_2$. We are assuming for now that $p_i$ and $p_j$ are
co-prime. Defining the vector 
\eqna
q=(q_1,q_2)\ ,\ \ \ \ 
&&q_1=\frac{1}{\sqrt{2}}\left(m_2p_ip_jL_2-im_3L_3\right)\ ,
\label{A.15}\\
&&q_2=\frac{1}{\sqrt{2}}\left(m_4L_4-im_5L_5\right)\ ,\nonumber
\eeqna
and using equation(\ref{A.13}) the condition (\ref{A.14}) reads
\eqn
g(z+q)=\a(q)\exp{\left(\frac{\pi}{2}H^{ij}(q,q)+\pi 
H^{ij}(z+z_0,q)\right)}g(z)\ ,
\label{A.16}
\eeqn
where $\a(q)$ is given by (\ref{4.34}) with $n^{ij}_{\ha\hb}$ defined
as in (\ref{A.9}). Using equation (\ref{4.33}) we see that $g(z)$ is
just a shifted $\Theta$-function, i.e. $g(z)=f(z+z_0)$. These
functions form a complex linear space of dimension
\eqn
|Pf(n^{ij})|=
|(p_jq_i-p_iq_j)(\bar{q}_i-\bar{q}_j)|=n^{ij}_L\bar{n}^{ij}_L\ .
\label{A.17}
\eeqn
Thus there are $n^{ij}_L\bar{n}^{ij}_L$ orthogonal ground state wave
functions obeying the boundary condition (\ref{A.14}).

\subsubsection*{Particle spectrum}

The particle excitations associated with the $a^i_{\a}$ and $c^i_m$
fields correspond to the usual wrapped D-brane spectrum
\cite{Polc1,Hash}. It is a $p$-fold degenerate spectrum with the momentum
along the $x^2$-direction quantised in units of $(p_iR_2)^{-1}$. We will
concentrate only on the massless torons associated with the
$b_{\a}^{ij}$ fields as these are the responsible for the entropy
formula (all the $d^{ij}_m$ excitations are massive as before).

We start by considering a given $a\bar{b}$ component of the $b^{ij}_{\ha}$
field that gives rise to the massless torons
\eqna
&&(b^{ij}_{z_k})_{a\bar{b}}=\sum_r (\chi^r_{_0})_{a\bar{b}}
\left( {\cal B}_{z_k}^{ij}\right)^r_0(x^\s)\ ,\ \ \ (k=1,2)\ ,
\nonumber\\\label{A.18}\\
&&(b^{ij}_{\bar{z}_k})_{a\bar{b}}=(b^{ij}_0)_{a\bar{b}}=
(b^{ij}_1)_{a\bar{b}}=0\ ,\nonumber
\eeqna
where as it will be explained below we have not written any
$U(p_i)\otimes U(\bar{p}_j)$ index in the conformal field 
$\left( {\cal B}_{z_k}^{ij}\right)^r_0(x^\s)$. The function 
$(\chi^r_{_0})_{a\bar{b}}$ satisfies the boundary conditions (\ref{A.14}).
Even though this function is defined for $x^2\in[0,p_ip_jL_2]$ the
field $\left( b_{z_k}^{ij}\right)_{a\bar{b}}$ takes values only in the
range $x^2\in[0,L_2]$. Using the first condition in equation
(\ref{A.11}) it may be seen that the conformal fields 
$\left( {\cal B}_{z_k}^{ij}\right)^r_0(x^\s)$ are in fact the same for all 
$\left( b_{z_k}^{ij}\right)_{a\bar{b}}$ components. The only
difference is the shift $z_0=\frac{1}{\sqrt{2}}(A_{a\bar{b}},0)$ in the
functions $(\chi^r_{_0})_{a\bar{b}}$. This is just a
consequence that the branes are wrapped and it means that the
$U(p_i)\otimes U(\bar{p}_j)$ field components
in $b_{\a}^{ij}$ are not independent. Thus as in the subsection 4.1.2
we have $4n^{ij}_L\bar{n}^{ij}_L$ massless bosonic torons
associated with the $b_{\a}^{ij}$ field.

The analysis is now entirely similar to the one presented before,
therefore it will not be repeated here. As a final remark, note that
we have assumed that $p_i$ and $p_j$ are co-prime. If this is not the
case, i.e. if $p_i=lp'_i$ and $p_j=lp'_j$ with $p'_i$ and $p'_j$
co-prime the ground state wave
function $(\chi_{_0})_{a\bar{b}}$ will obey the boundary conditions
(\ref{A.14}) with $p_ip_jL_2$ replaced by $lp'_ip'_jL_2$ and with
$n^{ij}_{23}=-n^{ij}_{32}$ replaced by $p'_jq_i-p'_iq_j$. The Landau
degeneracy for the $(\chi_{_0})_{a\bar{b}}$ functions will then be
\eqn
n'^{ij}_L\bar{n}^{ij}_L=|(p'_jq_i-p'_iq_j)(\bar{q}_i-\bar{q}_j)|\ .
\label{A.19}
\eeqn
However, when writing the fields $(b_{z_k}^{ij})_{a\bar{b}}$ as in
(\ref{A.18}) and imposing the first condition in equation (\ref{A.11})
we find that there are $l$ independent conformal fields 
$\left({\cal B}_{z_{k}}^{ij}\right)^{r,s}_0$ ($s=1,...,l$) for each
Landau level, i.e. each Landau level is itself degenerated. More
precisely, these independent conformal fields arise in the following
components of the fields
\eqn
(b_{z_k}^{ij})_{a\bar{b}}\ ,\ \ (b_{z_k}^{ij})_{a+1,\bar{b}}\ ,...,
\ \ (b_{z_k}^{ij})_{a+l-1,\bar{b}}\ .
\label{A.20}
\eeqn
This is in perfect agreement with the string theory picture described in
section 3 \cite{CostaPerry}.

\end{document}